\documentclass[floatfix,lengthcheck,sort&compress,reprint,aps,prl,superscriptaddress,nofootinbib,showpacs]{revtex4-1}

\usepackage[utf8]{inputenc}
\usepackage{amsmath}
\usepackage{amssymb}
\usepackage{dcolumn}
\usepackage{multirow}
\usepackage{graphicx}
\usepackage{dsfont}
\usepackage{mathrsfs}
\usepackage{units}
\usepackage{slashed}
\usepackage[usenames,dvipsnames]{xcolor}
\usepackage[colorlinks=true,linkcolor=blue,urlcolor=blue,citecolor=blue]{hyperref}

\begin{document}

\title{Vector mesons as dynamical resonances in the Bethe--Salpeter framework}

\author{Richard Williams}
\email{richard.williams@physik.uni-giessen.de}
\affiliation{Institut f\"ur Theoretische Physik, Justus-Liebig--Universit\"at Giessen, 35392 Giessen, Germany.}

\begin{abstract}
We present the first dynamical calculation of the $\rho$-meson as a resonance in the Dyson--Schwinger/Bethe--Salpeter framework by including explicit two-pion exchange in addition to the t-channel one-gluon exchange of rainbow-ladder in the interaction kernel. The width is determined from the imaginary part of the resonance pole and is generated by the singularity structure of the integrand, treated by means of suitable path deformations.  We find the resonant mass and width to be $640$~MeV and $100$~MeV respectively, corresponding to a coupling constant $g_{\rho\pi\pi}=5.7$; repulsive corrections beyond rainbow-ladder are known to bring these in alignment with experiment. Furthermore we present the dependence of these parameters on the pion mass and compare with results from lattice QCD.

\end{abstract}

  
\maketitle

{\it Introduction.}--- One of the long-standing goals of hadron physics is understanding the dynamics of the strong interaction. Below the strong-decay threshold, a number of bound-states \emph{e.g.} $\pi$, $K$, $D$, $B$, \emph{etc.} serve as a valuable playground for non-perturbative studies of QCD. Most hadrons, however, lie above these thresholds and manifest as resonances whose decay channels subsequently dominate in the decay width. Therefore any study is rendered incomplete when hadrons are described as stable bound-states instead of their ephemeral nature being embraced.

Many approaches have sought to describe the resonant structure of bound-states, e.g.~in constitutent quark models~\cite{Ricken:2003ua,Godfrey:1985xj,Kokoski:1985is,Feynman:1971wr}, lattice QCD~\cite{Aoki:2007rd,Gockeler:2008kc,Feng:2009ck,Feng:2010es,Aoki:2010hn,Frison:2010ws} and chiral perturbation theory~\cite{Hanhart:2008mx,Nebreda:2010wv}.
In Lattice QCD there has been significant progress recently in the determination of resonant properties of bound-states. Naively one expects no continuum of states due to finite volume---and hence no branch cut(s)---entailing that there is a single Riemann sheet. Consequently, unitarity dictates that all poles fall on the real-axis and one would conclude that resonances are inaccessible. However, decay properties may be inferred by performing a volume study of the spectrum~\cite{DeWitt:1956be,Luscher:1986pf,Luscher:1991cf}, with recent techniques such as distillation~\cite{Peardon:2009gh} contributing to improved signals and hence statistics.

In the continuum, where the Bethe--Salpeter equation (BSE) is formulated, resonances appear as poles in unphysical Riemann sheets (see e.g.~Refs.~\cite{PhysRev.121.1840,Pelaez:2015qba} for a discussion). However, they only arise dynamically when the interaction kernel explicitly incorporates decay processes, requiring truncations beyond that of rainbow-ladder~\cite{Williams:2015cvx,Binosi:2016rxz}. The exception to this is in systems where for structural reasons a resonance would be generated even with nothing more than one-gluon exchange, such as for tetraquarks e.g.~\cite{Eichmann:2015cra}; where however no resonance has been directly resolved yet due to limitations of the numerical methods thus far employed. To describe resonances \emph{dynamically} in the BSE we require two things: a (chiral-symmetry preserving) truncation that provides access to its decay channel; and a robust means by which the inevitable intermediate particle poles can be treated numerically.
It is the purpose of this letter to address both of these points for the explicit example of a $\rho$-meson. Generalization to \emph{e.g.} the scalar meson is straightforward and the approach may be applied to other hadrons such as baryons and---in particular---tetraquarks.

\begin{figure}[!t]
	\begin{center}
		{\centering\includegraphics[scale = 0.51]{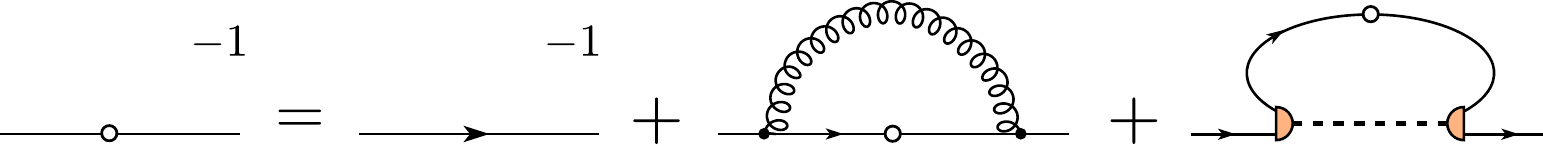}}\vspace{0.7cm}
		{\centering\includegraphics[scale = 0.51]{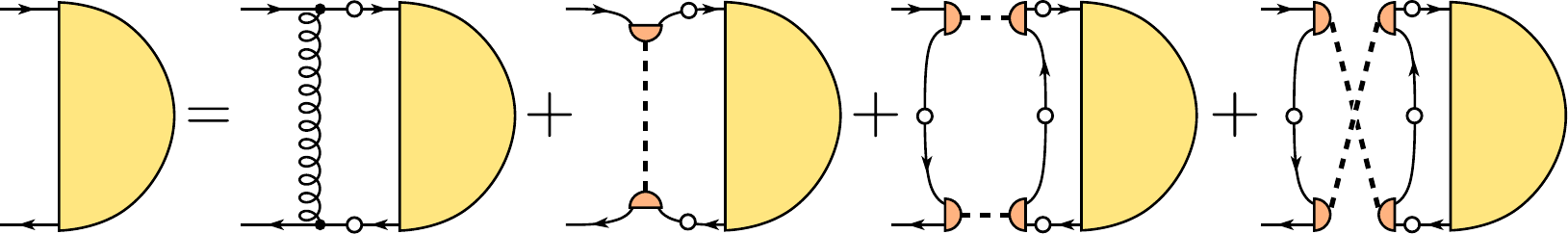}}
		\caption{\label{fig:dse_bse_rl_pi_pi_approx}(color online) Truncation of the quark DSE and meson BSE. Non-resonant binding between quarks (straight lines) is achieved by gluon exchange (wiggly lines), whilst the dominant resonance contribution is mediated by pions (small half-circles coupled by a dashed propagator). 
		}
	\end{center}
\end{figure}

{\it Framework.}--- We employ an approach based on the Dyson--Schwinger equations (DSEs), which have been successfully applied to a wide range of topics in particle physics and in particular QCD~\cite{Roberts:1994dr,Eichmann:2016yit}.  
In order to accommodate for decay channels, as inspired by Ref.~\cite{Watson:2004jq}, we employ the natural extension of the hadronic unquenching effects explored in Refs.~\cite{Fischer:2007ze,Fischer:2008sp,Fischer:2008wy} by complementing $t$-channel one-pion exchange with two-pion $s$- (and $u$- via crossing symmetry) channel exchange, see Fig.~\ref{fig:dse_bse_rl_pi_pi_approx}; detailed discussions of its origins can be found in Refs.~\cite{Watson:2004jq,Fischer:2007ze}. 
Such an extension has no deleterious impact on chiral symmetry, since the two-pion decay channel coupling to a pseudoscalar Goldstone boson is negligible as a result of both $P$ and $CP$.
Evidently, the dynamical coupling to the two-pion channel opens up the possibility for the parent state to decay, and will thus be realized as a resonance.

\begin{table*}[!t]
	\centering
	\caption{\label{tab:basis}Covariant basis for the pseudoscalar and vector mesons. The transverse objects are defined $\gamma_T^\rho = \gamma^\rho - \widehat{\slashed{P}}\widehat{P}^\rho$,
		and $p_T^\rho =      p^\rho -   p\cdot \widehat{P} \widehat{P}^\rho,$ with the hat  indicating normalization. Charge conjugation is enforced by including appropriate factors of $p\cdot P$.
	}
	\renewcommand{\arraystretch}{1.6}
	\begin{ruledtabular}
		\begin{tabular}{ll|llll}
			\multicolumn{2}{c|}{pseudoscalar}&\multicolumn{4}{c}{vector}\\
			\hline
			$\tau_1=\gamma^5$         
			& $\tau_3=\widehat{\slashed{P}}\gamma^5$          
			& $\tau^\rho_1=\gamma^\rho_T$
			& $\tau^\rho_3=i \widehat{p_T}^\rho$   
			& $\tau^\rho_5=3\widehat{p_T}^\rho\widehat{\slashed{p}_T} -\gamma^\rho_T$ 
			& $\tau^\rho_7=\gamma^\rho_T\widehat{\slashed{p}_T}-\widehat{p_T}^\rho$
			\\
			$\tau_2=\widehat{\slashed{p}_T}\gamma^5$    
			& $\tau_4=i\widehat{\slashed{p}_T}\widehat{\slashed{P}}\gamma^5$  
			& $\tau^\rho_2=\gamma^\rho_T\widehat{\slashed{P}}$   
			& $\tau^\rho_4=\widehat{p_T}^\rho \widehat{\slashed{P}}$ 
			& $\tau^\rho_6=(3\widehat{p_T}^\rho\widehat{\slashed{p}_T} -\gamma^\rho_T)\widehat{\slashed{P}}$
			& $\tau^\rho_8=i(\gamma^\rho_T\widehat{\slashed{p}_T}-\widehat{p_T}^\rho)\widehat{\slashed{P}}$  
			\\    
		\end{tabular}
	\end{ruledtabular}
\end{table*}

We start with the main constituent of our bound-state, the quark. Its inverse propagator $S^{-1}(p)=-i\slashed{p}A(p^2)+B(p^2)$ is described by the vector and scalar functions $A(p^2)$ and $B(p^2)$, obtained as solutions of the DSE
\begin{align}\label{eqn:quarkdse}
S_{ab}^{-1}(p) &= Z_2 S^{-1}_{0,ab}(p) + \Sigma_{ab}(p),
\end{align}
with the inverse bare quark-propagator $S^{-1}_0$ obtained by setting $A(p^2)=1$, and $B(p^2)$ to the bare quark mass $m_0$, and wavefunction renormalization is $Z_2$. The quark self-energy $\Sigma(p)$ is decomposed into a non-resonant and resonant contribution
\begin{align}\label{eqn:selfenergy}
\Sigma_{ab}(p)     &= \Sigma^{(0)}_{ab}(p) + \Sigma^{(1)}_{ab}(p),
\end{align}
where
\begin{align}
\Sigma^{(0)}_{ab}(p)&= \int_k \mathcal{K}^\mathrm{RL}_{ab;cd}(p,k,P) S_{dc}(k),\label{eqn:self0} \\
\Sigma^{(1)}_{ab}(p)&= \int_k \mathcal{M}_{ad;cb}(p,k,P) S_{dc}(k)\;,            \label{eqn:self1}
\end{align}
with $\int_k = \int d^4k/(2\pi)^4$ the loop-integration. The $\mathcal{K}^\mathrm{RL}$ is the rainbow-ladder kernel
\begin{align}\label{eqn:ladderkernel}
\mathcal{K}^\mathrm{RL}_{ab;cd}(p,k,P) = Z_2^2 C_F \left[ \gamma^\mu\right]_{ad}  g_s^2 D^{\mu\nu}(q) \left[ \gamma^\nu\right]_{cb},
\end{align}
where, $C_F=4/3$ is the color factor and $q=k-p$ the momentum of the gluon propagator $D^{\mu\nu}(q)$
; in this study the combination $g_s^2 D^{\mu\nu}(q)$ is replaced by the Maris-Tandy interaction~\cite{Maris:1999nt}, with its free parameter set to $\omega=0.4$~GeV.
Since we are calculating time-like properties of bound-states in Euclidean space, we require the quark propagator for complex momenta $p^2$ bounded by the parabola $k_\pm^2=k^2-M^2/4 \pm 2iM\sqrt{k^2}$, with $M$ the mass of the bound-state under investigation. Suitable techniques are discussed in the literature, see Refs.~\cite{Fischer:2008sp,Krassnigg:2009gd,Sanchis-Alepuz:2017jjd}.

The Bethe-Salpeter wavefunction $\chi^{(\rho)}$ can be constructed from the amplitude $\Gamma^{(\rho)}$ through
\begin{align}\label{eqn:bsewavefunction}
\chi^{(\rho)}(k,P) = S(k_+)\Gamma^{(\rho)}(k,P) S(k_-),
\end{align}
with $S(k_\pm)$ the incoming/outgoing quark propagators carrying momenta $k_\pm = k\pm P/2$. It satisfies the equation
\begin{align}\label{eqn:bethesalpeter}
\Gamma^{(\rho)}_{ab}(p,P) &= \int_k \mathcal{K}_{ab;cd}(p,k,P) \chi^{(\rho)}_{dc}(k,P),
\end{align}
where the Bethe-Salpeter kernel
\begin{align}\label{eqn:separationofkernels}
\mathcal{K}_{ab;cd}&=\mathcal{K}_{ab;cd}^{RL}+\mathcal{K}_{ab;cd}^{\pi,t}+\mathcal{K}_{ab;cd}^{\pi\pi,s}+\mathcal{K}_{ab;cd}^{\pi\pi,u},
\end{align}
is the sum of individual contributions derived from the self-energy $\Sigma$ through either a functional derivative of the quark self-energy~\cite{Munczek:1994zz} or via constraints imposed by the axial-vector Ward-Takahashi identity~\cite{Bender:1996bb,Fischer:2007ze,Chang:2009zb}. In this case, the resonant corrections $\Sigma^{(1)}$ and $\mathcal{K}^{\pi,t}$, $\mathcal{K}^{\pi\pi,s}$, and $\mathcal{K}^{\pi\pi,u}$ are derived under the assumption that at next-to-leading order in $1/N_c$, the sub-kernel $\mathcal{M}_{ad;cb}$ in~\eqref{eqn:self1} can be expressed by a ladder resummation. Accordingly~\cite{Watson:2004jq,Fischer:2007ze}
\begin{align}
\mathcal{M}_{ad;cb} &= \mathcal{K}^\mathrm{RL}_{ab;cd} + \int_k \mathcal{K}^\mathrm{RL}_{ab;mx} S_{xy} S_{nm} \mathcal{M}_{yn;cb},\label{eqn:Mresum} \\
          &\sim  \overline{\Gamma}^\pi_{ab}(p,q) D_\pi(q^2) {\Gamma}^\pi_{cd}(k,q),\label{eqn:Mresumapprox}
\end{align}
where in the second line the four-point function has been approximated by its dominant hadronic pole, $D_\pi(q^2)=\left(q^2+m_\pi^2\right)^{-1}$, and we have anticipated the association of this with a pseudoscalar pion of mass $m_\pi$. We drop the chiral symmetry preserving one-pion $t$-channel exchange $\mathcal{K}^{\pi,t}$ that has been reported elsewhere~\cite{Fischer:2008sp,Fischer:2008wy}, and focus on the kernels that facilitate the strong decay
\begin{align}
\mathcal{K}^{\pi\pi,s}_{ab;cd} &= \int_l \big[\Gamma^\pi S \Gamma^\pi\big]_{ab}D^\pi_+ D^\pi_- \big[\overline{\Gamma}^\pi S \overline{\Gamma}^\pi\big]_{cd},
\end{align}
where the integral is over the two-pion decay channel. The $u$-channel $\mathcal{K}^{\pi\pi,u}$ is trivially obtained by crossing symmetry, and $D_\pm^\pi=D^\pi(l_\pm^2)$ with $l=l\pm P/2$. 

The appearance of intermediate hadronic states  raises the possibility of simple poles entering the domain of integration when solving the BSE. These simple poles are intimately related to the decay, and it is their proper treatment and consequences that we now discuss.

{\it Calculation.}--- The amplitude $\Gamma^{(\rho)}$ and its wavefunction $\chi^{(\rho)}$ can be expanded in terms of the basis components $\tau^{(\rho)}_i$ (given in Table~\ref{tab:basis}), the construction of which for arbitrary spin is discussed in~\cite{Sanchis-Alepuz:2017jjd}
\begin{align}
\Gamma^{(\rho)} = \sum_{i} g_i \tau^{(\rho)}_i,\;\;\;\;
\chi^{(\rho)} = \sum_{i} h_i \tau^{(\rho)}_i,
\end{align}
with $g_i$ and $h_i$ scalar functions parameterizing the state. 
Orthogonality is defined through $\mathrm{Tr}\left[\overline{\tau}^{(\rho)}_i\tau_j^{(\rho)}\right] \propto \delta_{ij}$, where the conjugate basis element
\begin{align}\label{eqn:orthonormality}
\overline{\tau}_i(\widehat{p_T}, \widehat{P}) = (-1)^J \left[C^T\tau_i(-\widehat{p_T}, -\widehat{P})C\right]^T,
\end{align}
serves as a projector, with $J$ the total spin of the meson.

Our intent is to write the Bethe-Salpeter equation in the form $g_i = L_{ij} h_j$,
where $L_{ij}=\sum_A L^A_{ij}$ is factored into a sum of traces for each Bethe-Salpeter kernel $L^A_{ij}$ with $A$ indexing the kernels of~\eqref{eqn:separationofkernels}.
The $L^A_{ij}$ are obtained by projecting the $j$\textsuperscript{th}-component of $\mathcal{K}^A$ acting on  (the $h_j$ of) $\chi^\rho$ with the $i$\textsuperscript{th}-projector, defined in~Table~\ref{tab:basis} together with~\eqref{eqn:orthonormality}.

Specialising to the $rho$ meson and omitting flavor and color factors for brevity, the rainbow-ladder contribution yields
\begin{align}\label{eqn:nonresonantkerneltrace}
L_{ij}^\mathrm{RL} &=\int_k\mathrm{Tr}\left[\overline{\tau}^\rho_i(p,P) \gamma^\mu\tau^\rho_j(k,P)\gamma^\nu\right] D^{\mu\nu}(q),
\end{align}
where $q=k-p$ is the gluon momentum. 
For the $\pi\pi$ contribution in $s$-channel we have
\begin{align}\label{eqn:resonantkerneltrace}
L_{ij}^\mathrm{\pi\pi,s}&= \int_k\int_l \overline{J}^\rho_i(p,l,P)J^\rho_j(k,l,P) D^\pi_+D^\pi_-, 
\end{align}
where
\begin{align}
J^\rho_j(k,l,P)            &=	\mathrm{Tr}\left[\overline{\Gamma}_\pi\tau_j^\rho(k,P)\overline{\Gamma}_\pi S(k-l) \right], \\
\overline{J}^\rho_i(p,l,P)            &=	\mathrm{Tr}\left[\Gamma_\pi\overline{\tau}_j^\rho(p,P)\Gamma_\pi S(l-p) \right],
\end{align}
and similarly for the $u$-channel contribution.

If we examine the structure of~\eqref{eqn:resonantkerneltrace}, we see that the $\rho\pi\pi$ transition in impulse approximation 
\begin{align}
\Lambda^\rho_{\rho\pi\pi}(l,P) &=  \mathrm{Tr} \int_k \overline{\Gamma}^\pi S\Gamma^\rho S \overline{\Gamma}^\pi S, \\
                              &=2l^\rho_T F_{\rho\pi\pi}(l,P) + P^\rho G_{\rho\pi\pi}(l,P),
\end{align}
enters as a back-coupled sub-component, where $l$ is the relative momentum of the decay products transverse to parents' total momentum, $P$. The relevant form-factor is $F$ ($G$ is vanishing) which on-shell provides $g_{\rho\pi\pi}$~\cite{Jarecke:2002xd,Mader:2011zf}.

\begin{figure}[!t]
	\begin{center}
		\includegraphics[width=0.8\columnwidth]{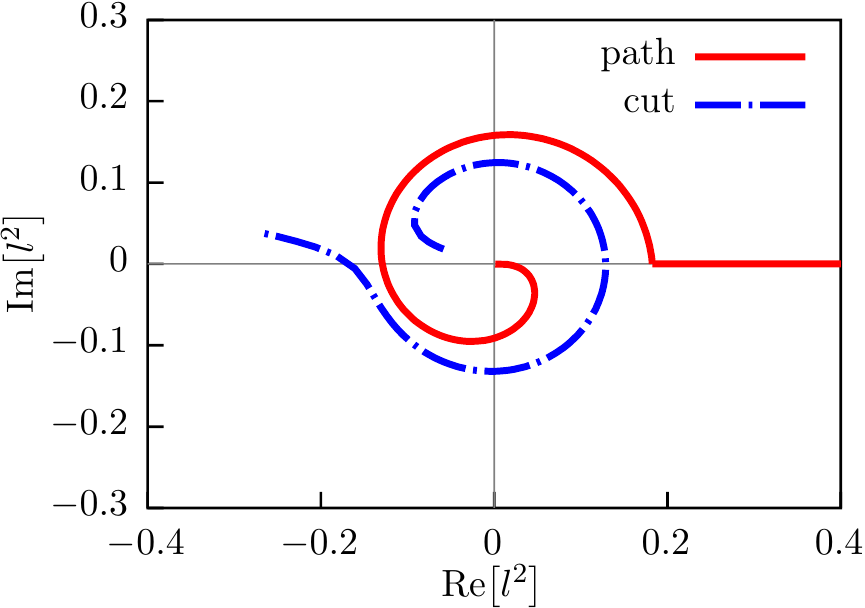}
		\caption{(color online) Singularity structure of the integrand due to the emergent pion poles, together with a suitable deformed path.}
		\label{fig:path_deformation}
	\end{center}
\end{figure}

When performing the numerical integral of~\eqref{eqn:resonantkerneltrace} in the iterative solution of~\eqref{eqn:bethesalpeter} we must consider the cut-structure of the integrand and treat it appropriately~\cite{Pichowsky:1999mu} by devising a suitable integration path~\cite{Maris:1995ns,Alkofer:2003jj,Weil:2017knt}. For equal mass particles (and with equal momentum partitioning) a single cut in $l^2\in\mathbb{C}$ is generated by the two propagator poles located at $l_\pm^2=(l\pm P/2)^2=-m_\pi^2$, given parametrically in terms of the cosine $z\in\left[-1,1\right]$ between $l$ and $P$
\begin{align}\label{eqn:cutequation}
l^2_\mathrm{cut} = -z \sqrt{t} + \sqrt{t(z^2-1)-m_\pi^2},
\end{align}
with $t = P^2/4$. This is shown in Fig.~\ref{fig:path_deformation}, together with an integration path that avoids the singularity structure.

\begin{figure}[!h]
	\begin{center}
		\includegraphics[width=0.7\columnwidth]{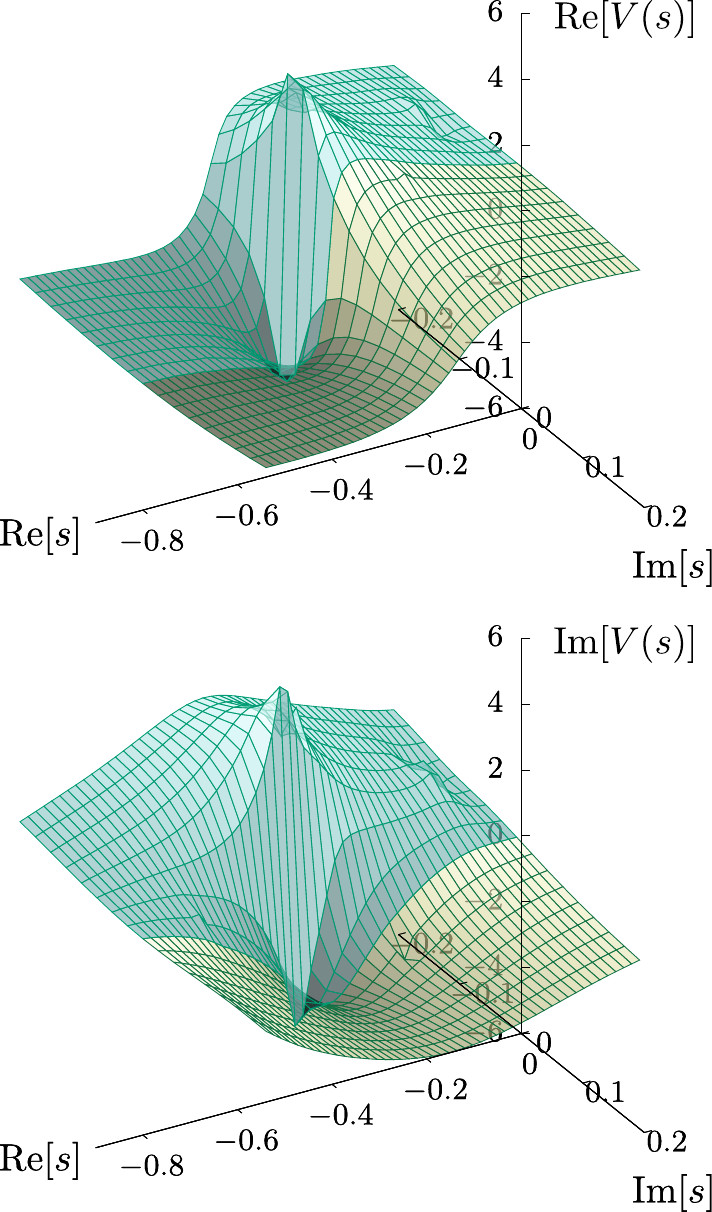}
		\caption{(color online) Real and imaginary parts of the inhomogeneous Bethe-Salpeter vertex, transformed by an inverse hyperbolic sine for better visibility. The upper half-plane is the first Riemann sheet, while the lower half-plane shows the continuation to the second Riemann sheet and the bound-state pole.}
		\label{fig:inhom_continuation}
	\end{center}
\end{figure}

\begin{figure*}[!t]
	\begin{center}
		\includegraphics[width=\textwidth]{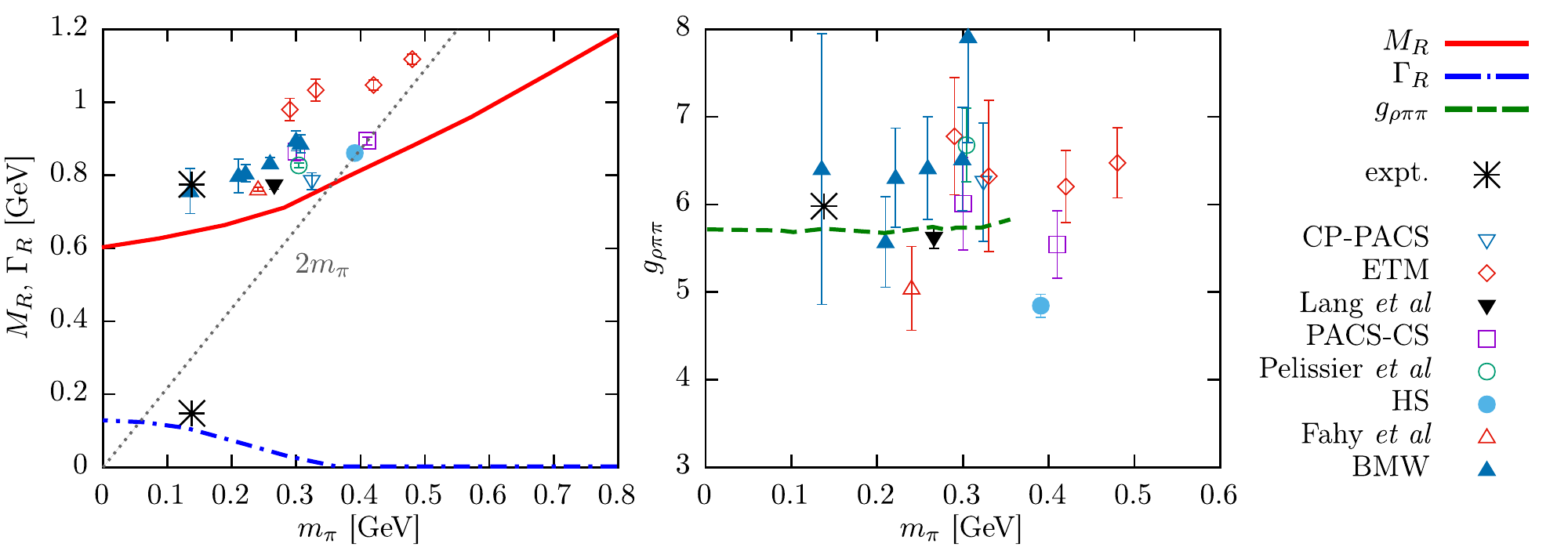}
		\caption{(color online) Calculated properties of the resonant vector meson as a function of the pion mass, compared to lattice data~\cite{Aoki:2007rd,Feng:2010es,Lang:2011mn,Aoki:2011yj,Pelissier:2012pi,Dudek:2012xn,Fahy:2014jxa,Metivet:2014bga}. The left panel shows the resonant mass $M_V$ and width $\Gamma_V$ as a function of the pseudoscalar mass $m_\pi$, compared with lattice data. The right panel displays the reconstructed strong coupling constant calculated through~\eqref{eqn:widthformula}.}
		\label{fig:mainresults}
	\end{center}
\end{figure*}

It remains to specify the off-shell behavior of the pion exchange in~\eqref{eqn:Mresumapprox}. We explored the scheme developed in Refs.~\cite{Eichmann:2008ef,Eichmann:2009zx}, and similarly applied to tetraquarks~\cite{Heupel:2012ua}, which retains all four components of the pion. However, the simpler prescription motivated from the observation that in the chiral limit the pion can be identified as the Goldstone mode of massless QCD
\begin{align}\label{eqn:boverfpi}
\Gamma_\pi(k,P) = \gamma_5 \frac{B(k^2)}{f_\pi},
\end{align}
where $f_\pi=92.4$~MeV is the leptonic decay constant and $B(k^2)$ is the scalar part of the quark propagator for vanishing quark mass,
provides for very similar results and simplifies both the presentation and discussion. Thus~\eqref{eqn:boverfpi} is employed for the remainder.

To solve the BSE we discretize the amplitudes radial and angular momenta and employ  quadrature for the remaining loop integrals, thus casting it as an eigenvalue equation
\begin{align}\label{eqn:eigenvalueequation}
\lambda\mathbf{x} = \mathbf{A}\,\mathbf{x},
\end{align}
by introducing the eigenvalue $\lambda=\lambda(P^2)$. For a resonance, $P^2$ will be complex valued and can be parameterized in terms of its resonant mass $M_R$ and width $\Gamma_R$ via $P^2 = -M_R^2+iM_R \Gamma_R$.
The eigenvalues are similarly complex-valued, with the solution found for discrete $P^2$ where $\lambda=1$. These solutions appear in complex conjugate pairs in the second Riemann sheet, necessitating analytic continuation across the branch cut. Alternatively, we can solve for the inhomogeneous BSE
\begin{align}\label{eqn:inhomogeneousequation}
\hat{\Gamma}_{ab}^\rho(p,P) = Z_2 \gamma^\rho + \int_k \mathcal{K}_{ab;cd}(p,k,P)\hat{\chi}^\rho_{dc}(k,P),
\end{align}
that complements~\eqref{eqn:bethesalpeter}, and directly resolve the resonance pole for $P^2$ complex after analytic continuation to the second Riemann sheet. In Fig.~\ref{fig:inhom_continuation} we plot $V(s)$ (with $s=P^2$) which is $\hat{\Gamma}(p,P)$ evaluated at $p^2=0$. Both methods, \eqref{eqn:eigenvalueequation} and~\eqref{eqn:inhomogeneousequation}, yield the same $P^2 = -M_R^2+iM_R \Gamma_R$.

\begin{table}[!b]
	\centering
	\caption{\label{tab:results}Masses and decay widths in GeV, together with the (dimensionless) strong coupling constant. The ${}^\dag$ indicates the result was extracted from~\eqref{eqn:widthformula}.}
	\setlength{\tabcolsep}{1em}
	\begin{ruledtabular}
		\begin{tabular}{l||c|c|c||c}
			&  $m_\pi$  &  $m_\rho$  &  $\Gamma_\rho$  &  $g_{\rho\pi\pi}$	\\
			\hline
			This work                              &  $0.14$  &  $0.64$   &  $0.10^{\phantom{^\dag}}$        &  $5.7^\dag$            \\
			DSE~\cite{Jarecke:2002xd,Mader:2011zf} &  $0.14$  &  $0.74$   &  $0.10^\dag$   &  $5.1^{\phantom{^\dag}}$            \\
			Experiment~\cite{Patrignani:2016xqp}   &  $0.14$  &  $0.78$   &  $0.15^{\phantom{^\dag}}$        &  $6.0^{\phantom{^\dag}}$            \\		
		\end{tabular}
	\end{ruledtabular}
\end{table}

{\it Results.}---
Assembling the pieces as described above and solving the BSE for a vector meson, with non-resonant binding described by a ladder exchange of gluons, and resonant effects pertaining to two-pion $s$- and $u$-channel exchange, we find
\begin{align}
P^2 = \left[i\left( 0.64 - \mathrm{i} 0.10/2\right)\right]^2 = -0.408 + i\, 0.065\,,
\end{align}
in GeV$^2$, for a physical pion mass. In the narrow width approximation we can reconstruct $g_{\rho\pi\pi}$ from $\left(M_R,\, \Gamma_R\right)$ via
\begin{align}\label{eqn:widthformula}
\Gamma_R = \frac{p^3 }{M_R^2}\frac{g_{\rho\pi\pi}^2}{6\pi},\;\;\;\;p=\sqrt{M_R^2/4-m_\pi^2}\;,
\end{align}
where $p$ is the relative momentum of the two pions at resonance.  The results are shown in Table~\ref{tab:results}, and compared with previous DSE studies working in the impulse approximation, together with experiment. The novel feature of the present study is that including the $\pi\pi$ decay channel leads dynamically to a negative shift of the resonant mass (compared to rainbow-ladder) by $100$~MeV, in addition to a width of comparable size. This leads to an improved value of $g_{\rho\pi\pi}=5.7$ as compared to the impulse approximation result of $5.1$. We  expect corrections beyond rainbow-ladder to be repulsive in the $\rho$ channel~\cite{Fischer:2009jm}, pushing up both the resonance mass and its width to be line with experiment.

This can be seen more clearly in the left-panel of Fig.~\ref{fig:mainresults} where we plot the resonant mass and width of the vector meson as a function of the pion mass viz. quark mass; both clearly underestimate the experimental value and lattice data, supporting the need to include the omitted non-resonant corrections.  The width clearly vanishes as the vector meson crosses the two-pion threshold. In the right-panel, we show the strong coupling $g_{\rho\pi\pi}$ as extracted from~\eqref{eqn:widthformula}. Its near-independence of the pion mass is clearly evident, as expected from chiral perturbation theory~\cite{Hanhart:2008mx,Hanhart:2014ssa}, with the cusp at threshold either indicative of small deviations beyond the strong phase-space dependence, or a result of details not captured in the Ansatz~\eqref{eqn:boverfpi}.

{\it Conclusion.}--- We have presented the first calculation of the $\rho$-meson as a resonance in the DSE/BSE, explicitly determining both real and imaginary parts of its pole, by including dynamically the dominant strong decay channel to two-pions, whilst maintaining the identification of the pion as a (pseudo)-Goldstone boson.  The numerical framework, wherein explicit poles in the integrand are carefully considered through path deformation, can be broadly applied to a large class of problems, notably the $f_0(500)$ in the tetraquark picture, as well as baryon resonances such as the $\Delta$.

%
\begin{acknowledgments}
We thank G.~Eichmann, C.~S.~Fischer, H.~Sanchis-Alepuz and P.~C.~Wallbott for useful discussions, and P.~Watson for input at an early stage of this work. 
This work has been supported by the Helmholtz International Center 
for FAIR within the LOEWE program of the State of Hesse, and the DFG Project No. FI 970/11-1.
\end{acknowledgments}

\bibliographystyle{apsrev4-1}
\bibliography{decay}

\end{document}